\documentclass[%
 reprint,
superscriptaddress,
 amsmath,amssymb,
aps,
]{revtex4-1}

\usepackage{graphicx}
\usepackage[utf8]{inputenc}
\usepackage{amssymb}
\usepackage{dcolumn}
\usepackage{bm}
\usepackage{xcolor}
\usepackage{todonotes}
\usepackage{hyperref}
\hypersetup{
    colorlinks=true,
    linkcolor=blue,
    filecolor=magenta,      
    urlcolor=red,
    citecolor=blue,
}

\usepackage{xcolor}

\usepackage{soul}

\begin{document}

\title{Mixed spin states for robust ferromagnetism in strained SrCoO$_3$ thin films}

\author{Xiquan Zheng}
\affiliation{International Center for Quantum Materials, School of Physics, Peking University, Beijing 100871, China}

\author{Nicholas B. Brookes}
\affiliation{ESRF-The European Synchrotron, 71 Avenue des Martyrs, CS 40220, F-38043 Grenoble, France}

\author{Flora Yakhou-Harris}
\affiliation{ESRF-The European Synchrotron, 71 Avenue des Martyrs, CS 40220, F-38043 Grenoble, France}

\author{Yingjie Lyu}
\author{Jianbing Zhang}
\affiliation{State Key Laboratory of Low Dimensional Quantum Physics, Department of Physics, Tsinghua University, Beijing 100084, China}

\author{Qian Xiao}
\affiliation{International Center for Quantum Materials, School of Physics, Peking University, Beijing 100871, China}

\author{Xinyi Jiang}
\affiliation{International Center for Quantum Materials, School of Physics, Peking University, Beijing 100871, China}

\author{Qingzheng Qiu}
\affiliation{International Center for Quantum Materials, School of Physics, Peking University, Beijing 100871, China}

\author{Qizhi Li}
\affiliation{International Center for Quantum Materials, School of Physics, Peking University, Beijing 100871, China}

\author{Shilong Zhang}
\affiliation{International Center for Quantum Materials, School of Physics, Peking University, Beijing 100871, China}

\author{Xinqiang Cai}
\affiliation{International Center for Quantum Materials, School of Physics, Peking University, Beijing 100871, China}

\author{Pu Yu}
\affiliation{State Key Laboratory of Low Dimensional Quantum Physics, Department of Physics, Tsinghua University, Beijing 100084, China}
\affiliation{Frontier Science Center for Quantum Information, Beijing 100084, China}

\author{Yi Lu}
\email{yilu@nju.edu.cn}
\affiliation{National Laboratory of Solid State Microstructures and Department of Physics, Nanjing University, Nanjing 210093, China}
\affiliation{Collaborative Innovation Center of Advanced Microstructures, Nanjing University, Nanjing 210093, China}

\author{Yingying Peng}
\email{yingying.peng@pku.edu.cn}
\affiliation{International Center for Quantum Materials, School of Physics, Peking University, Beijing 100871, China}
\affiliation{Collaborative Innovation Center of Quantum Matter, Beijing 100871, China}

\date{\today}

\begin{abstract}

Epitaxial strain in transition-metal oxides can induce dramatic changes in electronic and magnetic properties.  A recent study on the epitaxially strained SrCoO$_3$ thin films revealed persistent ferromagnetism even across a metal-insulator transition. This challenges the current theoretical predictions, and the nature of the local spin state underlying this robustness remains unresolved. Here, we employ high-resolution resonant inelastic x-ray scattering (RIXS) at the Co-$L_3$ edge to probe the spin states of strained SrCoO$_3$ thin films. Compared with CoO$_6$ cluster multiplet calculations, we identify a ground state composed of a mixed high- and low-spin configuration, distinct from the previously proposed intermediate-spin state. Our results demonstrate that the robustness of ferromagnetism arises from the interplay between this mixed spin state and the presence of ligand holes associated with negative charge transfer. These findings provide direct experimental evidence for a nontrivial magnetic ground state in SrCoO$_3$ and offer new pathways for designing robust ferromagnetic systems in correlated oxides.

\end{abstract}

\maketitle

Transition-metal oxides display a wide variety of emergent phenomena due to the delicate interplay between spin, charge, orbital, and lattice degrees of freedom. Epitaxial strain is a powerful method for tuning properties and can induce interesting phenomena such as the metal-insulator transition \cite{NNO_mit, LMO}, unconventional superconductivity \cite{LNO_1, LNO_2}, magnetic anisotropic \cite{SRO_CRO, CoCr2O4}, and spin state transition \cite{MNN, LCO_strain}. Among them, SrCoO$_{3}$ has drawn much attention because it exhibits a ferromagnetic metallic ground state that is highly sensitive to oxygen vacancies, doping, and lattice strain.  Density functional theory (DFT) calculations predict a transition from the metallic ferromagnetic (FM) state to an antiferromagnetic (AFM) metallic state at 2\% tensile strain \cite{DFT}. 
Experimentally, such a transition has indeed been observed in oxygen-deficient SrCoO$_{3-\delta}$ ($\delta <$ 0.2) thin films under tensile strain \cite{SCO_strain_FM_AFM}. However, isolating the pure effect of strain is challenging because tensile strain often induces oxygen vacancies in the unstable stoichiometric SrCoO$_{3}$ phase \cite{SCO_strain_O_vacancy}.

Recent advances in epitaxial growth and post-deposition ozone annealing techniques have enabled the synthesis of stoichiometric SrCoO$_3$ thin films under controlled strain \cite{XMCD_FM}. Intriguingly, these films exhibit a metal-insulator transition (MIT) under 2\% tensile strain, while retaining robust ferromagnetism below a Curie temperature (T$_c$) of ~250 K even under strains up to 3$\%$, as confirmed by x-ray magnetic circular dichroism (XMCD) measurements \cite{XMCD_FM}. 
This strain-insensitive magnetism contrasts sharply with theoretical predictions \cite{DFT}, suggesting a non-trivial interplay between electronic correlations and spin states. Resolving the local spin configuration is therefore essential to uncover the microscopic origin of robust ferromagnetism in SrCoO$_3$. Early studies proposed an intermediate-spin (IS) ground state \cite{IS_95, IS_HF}, while more recent dynamical mean-field theory (DMFT) calculations point to a complex admixture of spin and charge states involving fluctuating Co 3$d$ occupations \cite{spin_state_DMFT}. These contrasting models underscore the need for direct experimental evidence to establish the true spin-state character of SrCoO$_3$.

In this work, we employ high-resolution Co-$L_3$ edge resonant inelastic x-ray scattering (RIXS)  at beamline ID32 of ESRF \cite{ESRF_ID32} to directly probe the spin states of SrCoO$_3$ thin films ($\sim$ 30 nm) under varying tensile strains (1\%-3\%). By tracking the evolution of local $dd$ excitations across the ferromagnetic transition temperature (T$_c$), we observe a pronounced temperature-dependent change in spectral intensity, which cannot be attributed to magnetic or phonon excitations. Through detailed comparison with multiplet calculations, we identify the ground state as a mixture of high-spin and low-spin configurations, in contrast to the previously proposed intermediate-spin state \cite{IS_95}. Importantly, this mixed-spin state is stabilized by strong hybridization with ligand orbitals in a negative charge-transfer regime. The resulting spin-ligand entanglement leads to a moderate magnetic momentum and a robust ferromagnetic phase that persists across the strain-induced metal-insulator transition. Our results provide direct spectroscopic evidence linking mixed spin states and negative charge transfer to the stabilization of unconventional ferromagnetism in correlated cobalt oxides.

\begin{figure}[htbp]
\centering\includegraphics[width = \columnwidth]{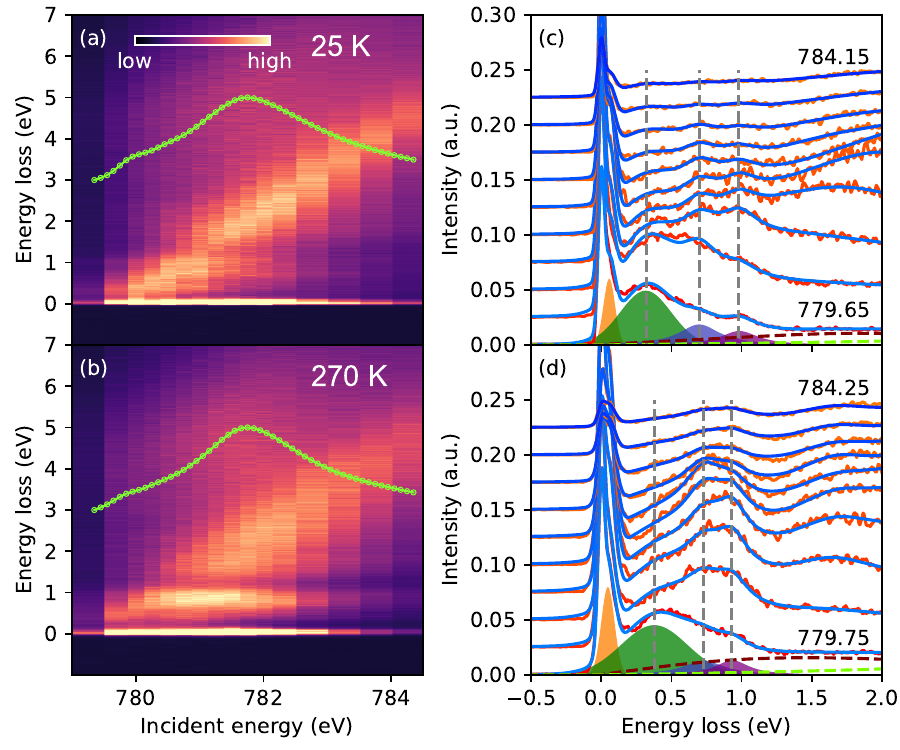}
\caption{\label{fig1} RIXS spectra of SCO$_3$/LAST thin films at \textbf{q} = (0.2, 0.2, 0.2) using $\pi$-polarized incident x-rays. \textbf{(a,b)} RIXS intensity
map versus energy loss and detuning energy across the Co $L_3$-edge at 25 K and 270 K, respectively. The green lines are Co $L_3$-edge x-ray absorption spectra with $\pi$-polarized incident x-rays collected at 25 K and 270 K, respectively, measured via total electron yield (TEY). \textbf{(c, d)} Selected RIXS spectra of
incident-energy detuning measurements at 25 K and 270 K with 0.5 eV incident energy interval, respectively. Each spectrum and fitting component were shifted vertically for clarity. Vertical lines indicate 
Raman-like features.
}
\end{figure}

In order to investigate the nature of the excitations in SrCoO$_3$ thin films, we performed RIXS measurements across the Co $L_3$ edge in 0.25~eV steps for SrCoO$_3$/La$_{0.3}$Sr$_{0.7}$Al$_{0.65}$Ta$_{0.35}$O$_3$ (SCO$_3$/LSAT) at 25~K and 270~K [Fig.~\ref{fig1}(a,b)]. The total energy resolution was $\sim$ 50~meV; additional experimental details are provided in Ref.~\cite{supplement}. Several Raman-like features are observed, characterized by energy-loss peaks that remain constant with varying incident photon energy. To quantitatively analyze these features, we fitted the RIXS spectra using Gaussian profiles and identified four distinct excitations. At 25~K, peaks are observed at approximately 0.06~eV, 0.32~eV, 0.70~eV, and 0.98~eV [Fig.~\ref{fig1}(c)]; at 270~K, they shift slightly to 0.06~eV, 0.36~eV, 0.73~eV, and 0.93~eV [Fig.~\ref{fig1}(d)]. The higher-energy excitations between 0.3 and 1.0~eV are consistent with local $d$--$d$ crystal field transitions \cite{LaCoO3,LaCoO3_dispersion}, whose energies exceed those of phonons ($<$ 90 meV) and magnetic excitations ($\leq$50 meV) typically observed in cobalt oxides \cite{CoTiO3, CoTiO3_2, 225_DFT, CaCoO2, LaCoO3_raman}.

\begin{figure}[htbp]
\centering\includegraphics[width = \columnwidth]{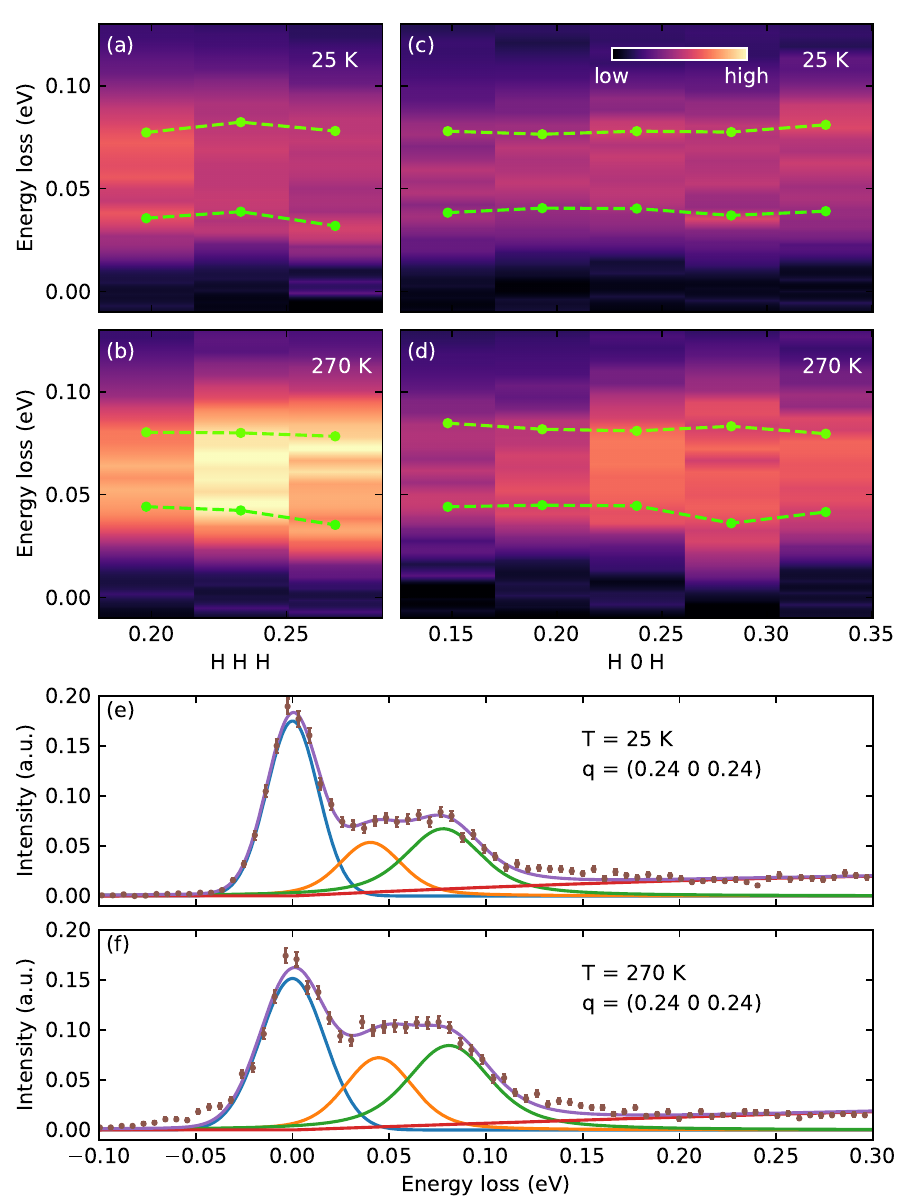}
\caption{\label{fig2}\textbf{(a,b,c,d)} Momentum dependence of low-energy features along [111] and [101] directions for SCO$_3$/LAST thin films at 25 K and 270 K. The spectra are normalized to fluorescence signals and the elastic peaks are subtracted.  The fit energies of two peaks are overlaid to the intensity map as green markers.  \textbf{(e,f)} The representative spectra and their fitting at momentum-transfer \textbf{q} = (0.24, 0, 0.24) at 25 K and 270 K, respectively. 
}
\end{figure}

To gain further insight into the Raman-like features observed below 0.1~eV, we performed high-resolution RIXS measurements with an overall energy resolution of $\sim$ 28~meV. Momentum-resolved RIXS spectra at the resonant energy of 781.7 eV were collected along the two high-symmetry directions, [101] and [111], as shown in Fig.~\ref{fig2}(a-d), with full experimental details provided in Ref.~\cite{supplement}. To reveal the low-energy excitations, the elastic peaks were subtracted from the spectra. The fits of the RIXS spectra are shown in Fig.~\ref{fig2}(e,f) and also in Supplementary Materials \cite{supplement}. Two distinct excitations are identified at energies of $\sim$ 45~meV and 80~meV. These features exhibit negligible momentum and temperature dependence along both measured directions, indicating their local origin. Interestingly, the intensities of both excitations increase by 20\% $\sim$ 70\% at 270~K compared to 25~K \cite{supplement}. This pronounced temperature dependence rules out a magnetic origin, as magnetic excitations typically weaken or vanish above the magnetic transition temperature. Moreover, the observed intensity changes are inconsistent with phonon excitations, which would yield only a modest enhancement of the anti-Stokes peak at elevated temperatures, as illustrated in Fig.~\ref{fig2}(f). Structural transitions are also excluded based on our x-ray diffraction measurements \cite{supplement}. Taken together, these observations suggest that the low-energy excitations are most likely associated with a change in the ground states, as discussed below.

\begin{figure}[htbp]
\centering\includegraphics[width = \columnwidth]{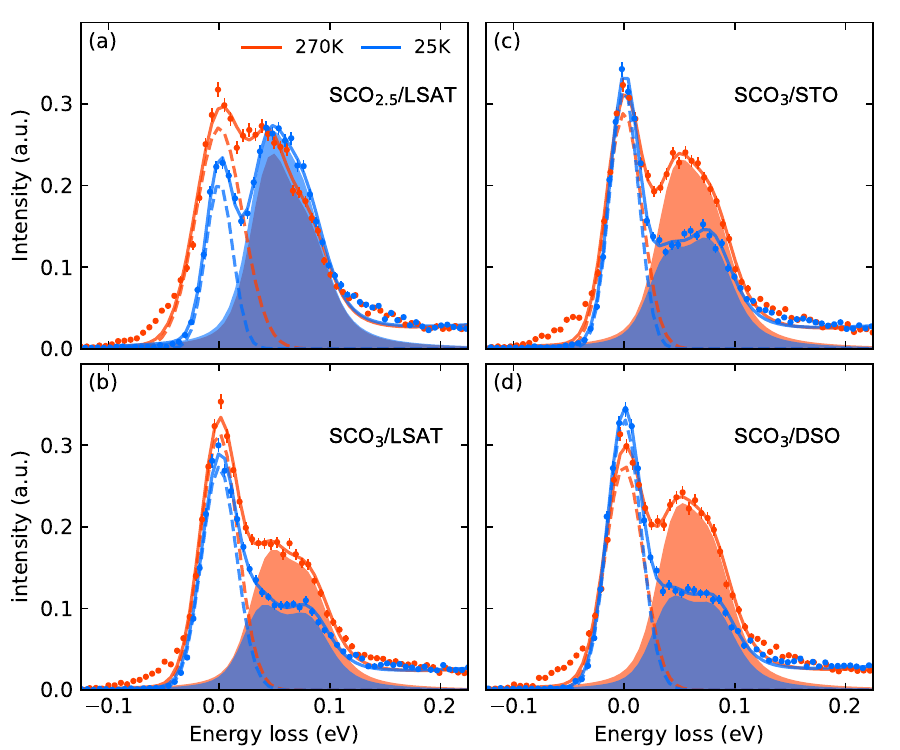}
\caption{\label{fig3}RIXS spectra at 25K and 270K for \textbf{(a)} SCO$_{2.5}$/LSAT, \textbf{(b)} SCO$_{3}$/LSAT, \textbf{(c)} SCO$_{3}$/STO, and \textbf{(d)} SCO$_{3}$/DSO with tensile strains of 1\%, 2\%, and 3\%, respectively. The momentum transfer is $\textbf{q}$ = (0.2, 0.2, 0.2). The elastic and low-energy excitations are marked with the dashed line and shaded areas. }
\end{figure}

To investigate the effect of epitaxial strain, we measured the SrCoO$_3$ thin films grown on La$_{0.3}$Sr$_{0.7}$Al$_{0.65}$Ta$_{0.35}$O$_3$ (LSAT), SrTiO$_3$ (STO), and DyScO$_3$ (DSO) substrates, which impose tensile strains of approximately 1\%, 2\%, and 3\%, respectively. For comparison, we also examined an as-grown SrCoO$_{2.5}$ (SCO$_{2.5}$) thin film. The growth procedures and structural characterization of all films are detailed in Ref.~\cite{XMCD_FM}. Figure~\ref{fig3} presents the RIXS spectra for these samples measured at 25~K and 270~K. In the SCO$_{2.5}$ film, the low-energy excitations change only slightly with temperature.  In contrast, all SCO$_3$ films exhibit a pronounced increase by more than 50\% in the intensity of low-energy excitations at 270~K relative to 25~K, while the excitation energies remain essentially constant across all strain levels. The high-energy $d$--$d$ excitations also show negligible variation with strain \cite{supplement}. These results indicate that the inelastic spectral features in SCO$_3$ are robust against epitaxial strain and undergo a similar temperature-driven transition in different strained thin films. This suggests that the underlying mechanism responsible for these excitations is an intrinsic property of SrCoO$_3$, largely insensitive to moderate tensile strain.

\begin{figure}[htbp]
\centering\includegraphics[width = \columnwidth]{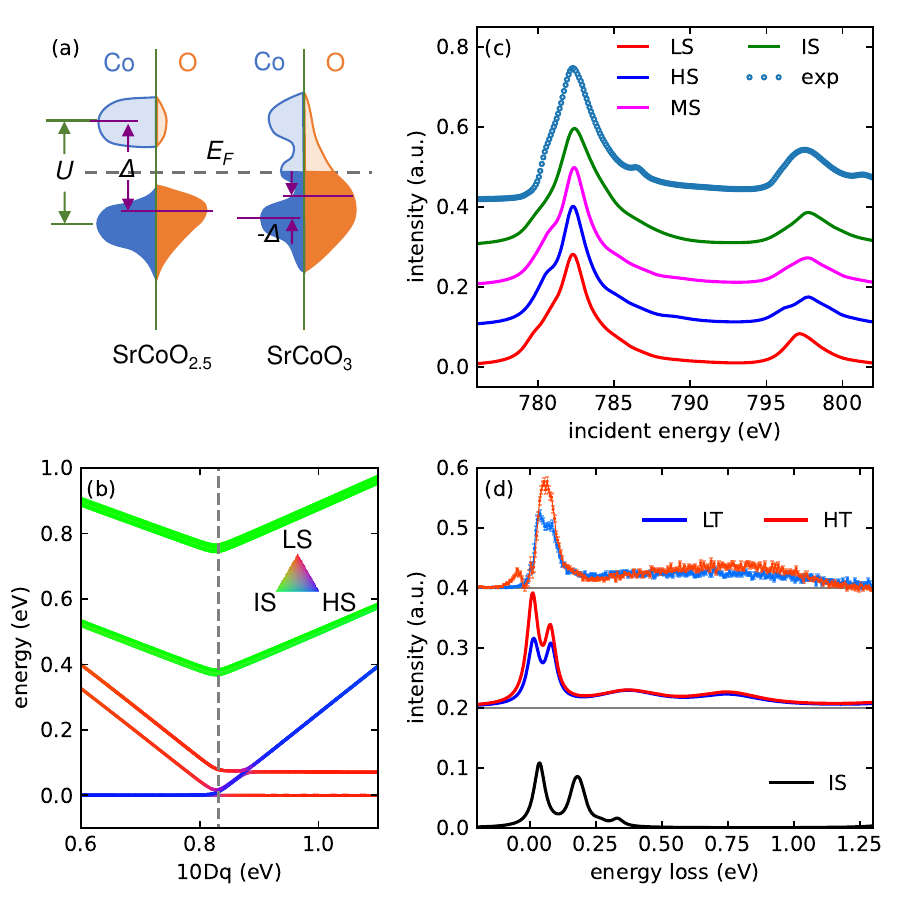}
\caption{\label{fig4}
\textbf{(a)} Illustration of energy band of positive charge-transfer (CT) energy material SrCoO$_{2.5}$ and negative CT energy material SrCoO$_{3}$. The CT energy is defined as the difference between the total energy with one ligand hole and with a full occupied ligand. As there are average 0.5 and 1 holes in ligand orbitals, the reference level is chosen at Fermi level and lower Hubbard band. \textbf{(b)} Calculated energy diagram of a CoO$_{6}$ cluster. The gray line indicates the crystal field energy used in the calculation. The color of energy levels corresponds with the distribution of total spin, as shown in the triangle. \textbf{(c)} Comparison of Co $L$-edge XAS spectrum collected in total electron yield (TEY) mode and $\sigma$ polarization to the calculated XAS with different spin states. \textbf{(d)} Calculated and experimental RIXS spectra with different configurations, after removal of their elastic component.}
\end{figure}

To interpret our experimental results, we performed multiplet calculations on a CoO$_6$ cluster within the framework of ligand field theory \cite{MLFT}, as detailed in the Supplementary Material \cite{supplement}. Assuming a nominal Co$^{4+}$ valence state, the cluster contains 15 electrons distributed over 20 spin-orbitals, allowing for three possible total spin configurations: high-spin (HS, $S = 5/2$), intermediate-spin (IS, $S = 3/2$), and low-spin (LS, $S = 1/2$). In modeling SrCoO$_3$, we adopted a negative charge transfer energy, in contrast to the positive charge transfer energy used for SrCoO$_{2.5}$ [Fig.~\ref{fig4}(a)]. By tuning the crystal field splitting parameter $10Dq$ from 0.6 to 1.1 eV, we constructed the energy-level diagram shown in Fig.~\ref{fig4}(b). This diagram indicates that the ground state favors either HS or LS character, depending on the interplay between crystal field strength and Hund's coupling \cite{LaCoO3_LS_to_IS, LCNO_IS}. When spin-orbital coupling (SOC) is included in the Co 3$d$ shell, a crossover regime emerges in which the ground state exhibits a mixed HS--LS character. Our calculated phase diagram differs from previous work reporting a stable IS state over a range of crystal field values \cite{IS_95}, primarily due to stronger Co-O hybridization in our model. Specifically, we used $V_{e_g} = 3.25$ eV and $V_{t_{2g}} = 1.8$ eV, almost 67\% higher than the values reported in Ref.~\cite{IS_95}. These hybridization strengths are supported by independent DFT studies \cite{DFT_MLFT, DFT_W_MLFT} and are consistent with values found in other cobaltates featuring a similar CoO$_6$ cluster geometry \cite{Co3O4}.

Figure~\ref{fig4}(c) shows the calculated XAS spectra for HS, LS, IS, and mixed HS-LS (MS) states, alongside the experimental data. The Co $L_3$ edge features a main peak with a lower-energy shoulder, and the $L_2$ edge displays a broad single peak. Although prior work suggested good agreement with the IS state \cite{IS_95}, the subtle spectral differences make it difficult to reach a definitive conclusion based on XAS alone. To resolve this, we computed the RIXS spectra, shown in Fig.\ref{fig4}(d), using the crystal field parameter indicated by the dashed line in Fig.\ref{fig4}(b). The red and blue curves correspond to low and high temperatures, respectively, with identical parameters except for a small exchange field applied at low temperature to simulate ferromagnetic spin alignment. Both spectra show similar excitation energies but differing intensities. The pronounced enhancement below 0.1 eV arises from the transitions between LS states split by SOC, matching the experimental low-energy feature. Our calculations on RIXS spectra reveal significant intensity for the spin-flip processes including from LS to IS and HS to IS states due to the strong SOC on the Co 2p core shell but negligible RIXS intensity for HS to LS excitations \cite{supplement}. For comparison, the calculated RIXS spectrum for the IS states with parameters from Ref.~\cite{IS_95} (black line) shows evident peaks at $\sim$ 0.2 and 0.3 eV, but does not show any significant intensity between 0.4 and 1.0 eV, contrary to the experiment. Moreover, a purely IS ground state would not exhibit the observed temperature-dependent d-d excitations. These results strongly rule out an IS ground state in SrCoO$_3$ and support a mixed HS-LS configuration.

Previous DMFT calculations~\cite{spin_state_DMFT} identified a distribution of spin states on Co $3d$ orbitals in SrCoO$_3$, with the $d^6$ high-spin (HS, $S = 2$) configuration dominating approximately 40\%, followed by the intermediate-spin (IS, $S = 1$) state at approximately 20\%. Our cluster calculations, which consider the total spin of the full CoO$_6$ cluster, including the contributions of the ligands, provide an extended perspective. We find a broad distribution of charge states, with $d^5$, $d^6$, $d^7$, and $d^8$ configurations occurring with probabilities of 22\%, 52\%, 23\%, and 2\%, respectively. Within the dominant $d^6$ component, the total-spin probabilities are 33\% for $S = 2$, 15\% for $S = 1$, and 4\% for $S = 0$. These results are consistent with the DMFT trends but underscore the critical role of ligand hybridization and spin-charge entanglement in stabilizing the mixed-spin ground state.

The metal-insulator transition in SrCoO$_3$ occurs without any accompanying magnetic phase transition, highlighting the unusual decoupling between charge and spin degrees of freedom in this system. In contrast, the related positive charge transfer compound SrCoO$_{2.5}$ exhibits robust G-type AFM order up to 570 K  \cite{SCO225_magnetic}. Similarly, in iron oxides, the positive charge transfer compound Sr$_2$Fe$_2$O$_5$ exhibits AFM order up to $\sim$690 K \cite{Sr2FeO4_PND}. These results suggest a critical role for charge transfer energy in determining the magnetic ground state and its stability. To elucidate the origin of magnetic coupling in SrCoO$_3$, we constructed a model including 3d orbitals of two cobalt sites and their corresponding ligand orbitals, allowing us to compute the ground state evolution as a function of charge transfer energy \cite{bonddisproportionation}. Our calculations reveal that FM exchange coupling emerges over a range of negative charge-transfer energies \cite{supplement}.  Furthermore, with the two-site calculation, we find an extended parameter space within this FM regime that stabilizes the mixed spin states compared to that predicted by single-site ligand field calculations. This result suggests the inherent stability of mixed spin configurations in SrCoO$_3$, underscores the importance of negative charge transfer and persistent ligand holes in mediating ferromagnetic interactions.

Our RIXS measurements reveal that the local electronic structure in SrCoO$_3$ thin films remains largely insensitive to epitaxial strain, consistent with previous XAS and XMCD studies \cite{XMCD_FM}. This robustness reflects the dominant role of indirect exchange interactions mediated by ligand holes, which preserve the alignment of Co 3$d$ magnetic moments even as strain suppresses itinerant carriers. Although the mixed-spin state is relatively unaffected by strain, alternative tuning parameters, such as chemical doping or electrostatic gating, can shift the high-spin-low-spin balance (HS-LS) and stabilize emergent magnetic phases. For example, ionic liquid gating enables a reversible transformation between SrCoO$_{2.5}$ and SrCoO$_3$ by modulating oxygen vacancies \cite{Lu2017,ionic_liquid_gating}. On the other hand, substituting Sr with La dopes electrons and stabilizes the low-spin ($S = 0$) ground state at low temperatures \cite{LaCoO3, LaCoO3_LS_to_IS_eels}. The combined control of oxygen stoichiometry and La substitution thus offers a promising avenue for tailoring magnetic ground states, as has been demonstrated in recent work \cite{LaCoO3_strain_FM_eels}. 

In summary, by combining high-resolution RIXS measurements with multiplet cluster calculations, we have identified the ground state of strained SrCoO$_3$ thin films as a mixture of HS and LS configurations. This unique mixed-spin ground state, stabilized by strong Co-O hybridization and negative charge transfer energy, provides a consistent explanation for the persistence of ferromagnetism across the strain-induced metal-insulator transition. Our results challenge the previously proposed intermediate-spin scenario and offer an experimentally validated picture of spin-state mixing in cobalt-based oxides. Importantly, the robustness of ferromagnetism in SrCoO$_3$ is not merely a consequence of strain-driven distortions, but arises from the intrinsic interplay between spin-state fluctuations and ligand hole dynamics. This insight opens a new route toward engineering strain-resilient magnetic states in correlated oxides by exploiting spin-ligand entanglement. The ability to tune the ratio of high- and low-spin components, potentially via electrostatic gating or chemical substitution, may enable new functional magnetic phases and device paradigms based on mixed-spin physics.

Y.Y.P. is grateful for financial support from the National Natural Science Foundation of China (Grants No. 12374143 and No. 11974029), the Ministry of Science and Technology of China (Grants No. 2019YFA0308401 and No. 2021YFA1401903), and Beijing Natural Science Foundation (Grant No. JQ24001). Y.L. acknowledges the financial support from the Ministry of Science and Technology of China (Grant No. 2022YFA1403000) and the National Natural Science Foundation of China (Grant No. 12274207). P.Y. acknowledges the financial support from the National Natural Science Foundation of China (Grants No. 52025024 and No. 12421004) and National Key R\&D program of China (Grant No. 2023YFA1406400). The experimental data were collected at the beam
line ID32 of the European Synchrotron (ESRF) in Grenoble
(F) using the ERIXS spectrometer designed jointly by
the ESRF and Politecnico di Milano.

\end{document}